\def\Missing#1#2{
         \ifmmode
              {#1}\kern-0.6em\lower-.1ex\hbox{/}_{#2} 
         \else
             ${#1}\kern-0.6em\lower-.1ex\hbox{/}_{#2}$
         \fi}
\def\met{\mbox{${\hbox{$E$\kern-0.6em\lower-.1ex\hbox{/}}}_T$}} 
\def\D0{D\O}                            
\def\d0draft{}
\def\err#1#2#3 {{\it Erratum} {\bf#1},{\ #2} (19#3)}
\def\ib#1#2#3 {{\it ibid.} {\bf#1},{\ #2} (19#3)}
\def\nc#1#2#3 {Nuovo Cim. {\bf#1} ,#2(19#3)}
\def\nim#1#2#3 {Nucl. Instr. Meth. {\bf#1},{\ #2} (19#3)}
\def\np#1#2#3 {Nucl. Phys. {\bf#1},{\ #2} (19#3)}
\def\pl#1#2#3 {Phys. Lett. {\bf#1},{\ #2} (19#3)}
\def\prev#1#2#3 {Phys. Rev. {\bf#1},{\ #2} (19#3)}
\def\prl#1#2#3 {Phys. Rev. Lett. {\bf#1},{\ #2} (19#3)}
\def\rmp#1#2#3 {Rev. Mod. Phys. {\bf#1},{\ #2} (19#3)}
\def\zp#1#2#3 {Zeit. Phys. {\bf#1},{\ #2} (19#3)}

\input{epsf}
\documentclass[epj]{svjour}

\usepackage{graphics}
\begin{document}
\title{On fitting the Pareto-Levy distribution to stock market index data: selecting a suitable cutoff value.}

\author{H.F. Coronel-Brizio\  \thanks{\emph{e-mail:} hcoronel@uv.mx} 
and\ A.R. Hernandez-Montoya\ \thanks{\emph{e-mail:} alhernandez@uv.mx}%
}

\institute{Facultad de F\'{\i}sica e Inteligencia Artificial.
Universidad Veracruzana, Apdo. Postal 475. Xalapa, Veracruz. M\'{e}xico.}
\date{Received: date / Revised version: date}
\abstract{The so-called Pareto-Levy or power-law distribution has been successfully used 
as a model to describe probabilities associated to extreme variations of worldwide stock markets indexes data and it has the form $Pr(X>x) \simeq x^{-\alpha}$ for
$\gamma<x<\infty$. The selection of the threshold parameter $\gamma$ from empirical data and consequently,
the determination of the exponent $\alpha$, is often is done by using a simple graphical method
based on a log-log scale, where a power-law probability plot shows a straight line with slope equal to the exponent of the power-law distribution.
This procedure can be  considered subjective, particularly with regard to the
choice of the threshold or cutoff parameter $\gamma$. In this work is presented  a more objective procedure, based on a statistical measure of discrepancy between the empirical and the Pareto-Levy distribution.
The technique is illustrated for data sets from the New York Stock Exchange
Index and the Mexican Stock Market Index (IPC). 
\\
} 
\authorrunning{H.F. Coronel-Brizio and A.R. Hernandez-Montoya}%

\titlerunning{On fitting the Pareto-Levy distribution}

\maketitle
\section{Introduction}
\label{intro}
\vspace*{-.3cm}
\noindent
The Power-law distribution is present in a great scope of physical (phase transitions, nonlinear dynamics and disordered systems) \cite{physics1,physics2}, financial (stocks prices and indexes variations,
\noindent
volumes, volatility 
decay distributions)~\cite{plerou,GOPI98,GOPI99,lux,volumes1,model2,voldecay2,volpar}, and 
other kinds of social phenomena (the World Wide Web and Internet router links,
sexual contact networks, growth of cities, reference networks in scientific journals, 
University entrance examinations, and traffic penalties distributions)
\cite{www,router,sexual,cities,citas,education,multas}. 
All these systems share the property of complexity and are driven by collective mechanisms of which signature is the power-law 
distribution.

\noindent
Studying variations of financial data is important in order to understand  the stochastic process that drives them and also for practical purposes related to investment and risk management.

\noindent
In the analysis of stock market indexes variations, many observables are used\cite{mantegnabook}. In this work we have chosen the returns series. 
\noindent
Briefly reviewing the definition of series, if in general,  $X(t)$ denotes 
the value of a particular index at time $t$, its return series is defined as  
$S(t)=\log X(t+\Delta t)-\log X(t)$; that is, as the logarithmic changes in the values
 of the index for a certain interval of time $”\Delta t$, which can be studied within a few seconds to a many days range.

\noindent
It has been reported  in several empirical studies, that in order to describe the probabilities of extreme returns variations, the Pareto-Levy distribution is an useful model to compute 
probabilities. However, when fitting the power-law to empirical data, the 
choice of the threshold or cut-off parameter of the tail distribution does not 
seem to follow an objective procedure. Usually, its fitted value is obtained 
by judging the degree of linearity in a log-log plot involving the empirical 
and theoretical distributions, even more, recently, well founded studies have 
criticize the reliability of this  geometrical method~\cite{Goldstein,weron} . 

\begin{figure}[htbh]
\resizebox{0.5\textwidth}{!}{%
\includegraphics{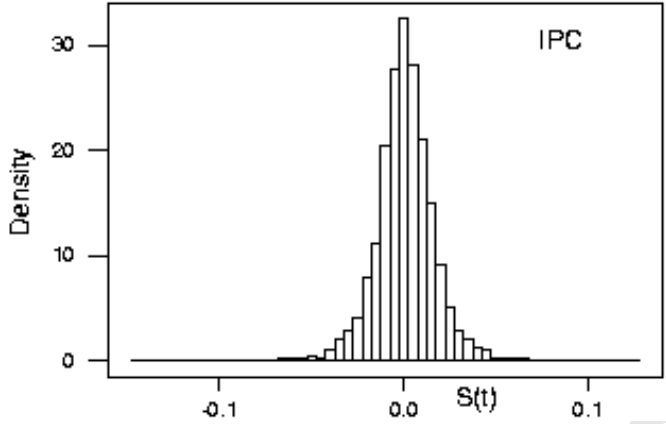}}
\caption{Density histogram for daily logarithmic differences
of the Mexican IPC index, from April 19 1990 to September 17 2004}
\label{histoipc}
\end{figure}

\begin{figure}[!htb]
\resizebox{0.5\textwidth}{!}{%
\includegraphics{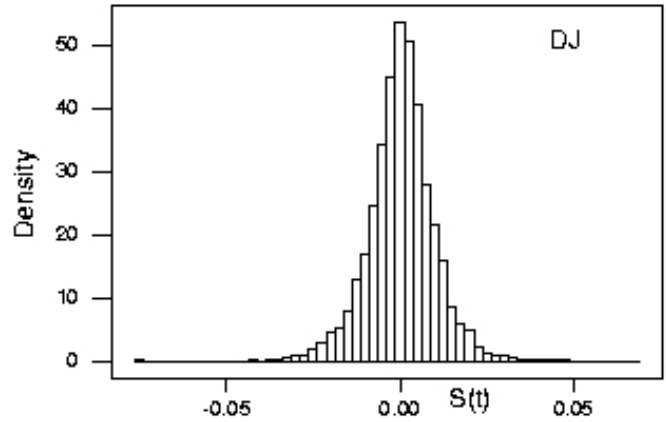}}
\caption{Density histogram for daily logarithmic differences
of the Dow Jones index, from April 19 1990 to September 17 2004}
\label{histodj}
\end{figure}

\noindent
Independently of these studies just cited above, we propose in this work, a formal procedure to improve the 
quality of the log-log plot fit based  on measures of discrepancy between the 
empirical and  theoretical distribution  functions.
\noindent
In order to illustrate our technique, we study the daily returns distribution of both, an emergent and a well developed stock 
markets: the Mexican stock market index IPC~\footnote{\'Indice de Precios y Cotizaciones, which means
Prices and Quotations index} and the American DJIA~\footnote{Dow Jones Industrial Average Index}. Figures \ref{histoipc} and \ref{histodj} show daily returns distributions respectively  for these financial markets.

\section{Choosing the value of the threshold parameter}
\vspace*{-.3cm}

\noindent
Let $S(s)$ be an absolutely continuous random variable and let us assume that there exists 
a value $\gamma$, such that, for $s>\gamma$, $S$ follows a Pareto-Levy distribution
with parameters $\alpha>0$ and $\gamma$; that is: 

\noindent
When $s>\gamma$,

\begin{equation}\label{pareto1}
F_{0}(s)=1-\left(\frac{\gamma}{s}\right)^{\alpha}
\end{equation}
\noindent
If the value $\gamma$ was known, we could compute a measure of fit
of the Pareto-levy model using, for example, quantities within the so-called
quadratic statistics:

\[
Q_n  = \int\limits_{ - \infty }^\infty  {\left[ {F_n  - F_0 } \right]^2 \psi dF_0 } 
\]

\noindent
where $F_{n}$ denotes the empirical distribution function and $\psi$ is a weighting
function; for example:
\begin{itemize}

\item  For $\psi(z)=1$, $Q_{n}$ becomes Watson's $W^{2}$ statistic.\\

\item For $\psi(z)=\left\{ {F_0 (z)\left[ {1 - F_0 (z)} \right]} \right\}^{ - 1}$  $Q_{n}$ it is the well known Anderson-Darling $A^{2}$ statistic.\\
\end{itemize}
\noindent
Focusing in the former, if we denote by $A^{2}(\gamma)$ the computed value of the $A^{2}$ statistic
for a given value $\gamma$, the fitted value of $\gamma$ can be chosen  as the value 
which minimizes $A^{2}(\gamma)$.
\noindent
In the next section the computational details will be described; however, for a more
detailed treatment of statistics based on the empirical distribution function, the
interested reader is referred to \cite{stephens}.

\section{Computing formulas}
\vspace*{-.3cm}

\noindent
Let $s_{(1)}\leq s_{(2)} \leq \ldots \leq s_{(n(\gamma))}$ denote the ordered values of the
observed series $s(t_{1})$, $s(t_{2}),\ldots$, $s(t_{n(\gamma)})$, where $s(t_{i})\ge \gamma$ for
$i=1,\ldots,n(\gamma)$ and $n(\gamma)$ denotes the number of remaining observations in the
sample which are greater than or equal a given admissible value of $\gamma$. Then the procedure 
steps of our method, can be enumerated as follows:

\begin{enumerate}
\item  Estimate the shape parameter $\alpha(\gamma)$ by 
\[
\alpha \left( \gamma  \right) = \left[ {\frac{1}{n(\gamma)}\sum\limits_{i = 1}^{n(\gamma )}
{\log \left( {\frac{{s_i }}{\gamma }} \right)} } \right]^{ - 1} 
\]
\item For $i=1,\ldots,n(\gamma)$, compute the quantities
\[
z_{\left( i \right)}  = 1 - \left( {\frac{\gamma }{{s_{\left( i \right)} }}}
\right)^{\alpha \left( \gamma  \right)}
\]
\item Compute the value of the Anderson-Darling statistic using

\begin{eqnarray}
A^{2}(\gamma)&=&{ - n(\gamma ) - \frac{1}{{n(\gamma )}}\sum\limits_{i = 1}^{n(\gamma )}
 {\left( {2i - 1} \right)\left[ {\log z_{\left( i \right)} } \right.} } \nonumber \\
    &+& {\left. {\log \left\{ {1 - z_{\left( {n - i + 1} \right)} } \right\}} \right]} \nonumber
\end{eqnarray}
\end{enumerate}
\noindent
Starting with, say, $\gamma_{1}=s_{(1)}$ in the complete sample, a
sequence $\gamma_{1},\ldots,\gamma_{r}$ can be constructed, to a desired
accuracy, to produce the sequence of values $A^{2}(\gamma_{1}),\ldots,A^{2}(\gamma_{r}).$

\noindent
A plot of $\gamma_{r}$ versus $A^{2}(\gamma_{r})$ will be useful,
as it will be shown in the next section, for finding the value of $\gamma$ which minimizes
the value of $A^{2}.$

\begin{figure}[t]
\resizebox{0.50\textwidth}{!}{%
\includegraphics{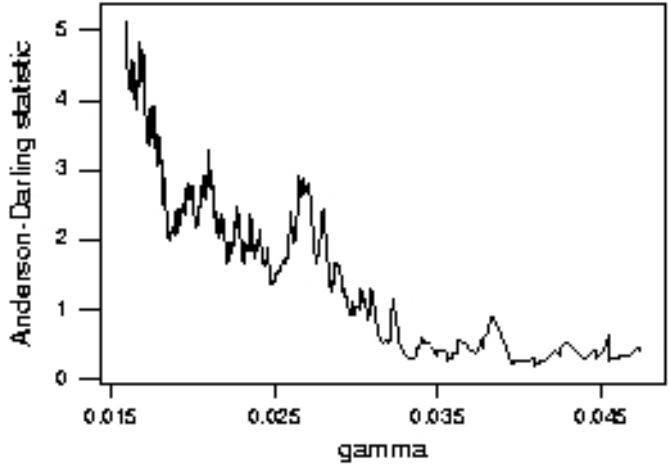}}
\caption{Anderson-Darling $A^2$ statistic versus selected values of the
threshold parameter $\gamma$, corresponding to the positive values of the series
$S(t)$, computed from the IPC index data. The minimum value, 0.16, is
attained for $\gamma=0.0395$}
\label{maxa2ipcp}
\end{figure}
\section{Data Analysis}
\vspace*{-.3cm}

\noindent
In order to illustrate the technique, two data sets were
analyzed. For both data sets, the daily returns series $S(t)$ was
constructed.
 The first one, consists of daily values
of the Mexican stock market index (IPC), covering from 
April 19, 1990 to September 17, 2004. The $S(t)$ series
which we will denote here as $S_{IPC}(t)$, had 3608
values from which 1877 positive, and 1723 negative values,
were used for the analysis.
 The second data set consists of daily values
of the Dow Jones index recorded from April 19, 1990
to September 17, 2004. The $S(t)$ series constructed
from this data, denoted here by $S_{DJ}(t)$, had 3633
values. Here the analysis was based on 1899 and 1723
positive and negative values, respectively.
\noindent
Data bases for the IPC and DJIA were downloaded from \cite{ipc_data} and \cite{dj_data} respectively.

\vspace{.08cm}
\noindent
Using the procedure described in the previous section,
the Pareto distribution was fitted to the positive and
negative tails of the distributions of $S_{IPC}(t)$ and
$S_{DJ}(t)$ varying the value of the parameter $\gamma$
over the ordered sample values. In each case, the Anderson-Darling
statistic, was used as a goodness-of-fit criterion.
It must be remarked that for the analysis of the negative tails,
the values $-S(t)$ where used.

\begin{figure}[htbp]
\resizebox{0.50\textwidth}{!}{%
\includegraphics{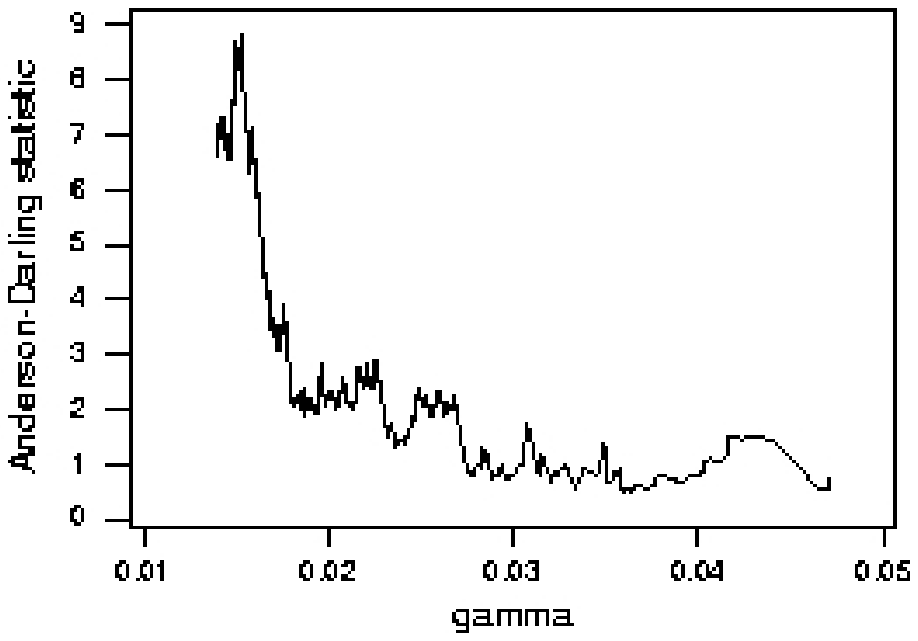}}
\caption{Anderson-Darling $A^2$ statistic versus selected values of the
threshold parameter $\gamma$, corresponding to the negative values of the series $S(t)$, computed from the IPC index data.The minimum value, 0.50, is
attained for $\gamma=0.036$}
\label{maxa2ipcn}
\end{figure}

\begin{figure}[htbp]
\resizebox{0.50\textwidth}{!}{%
\includegraphics{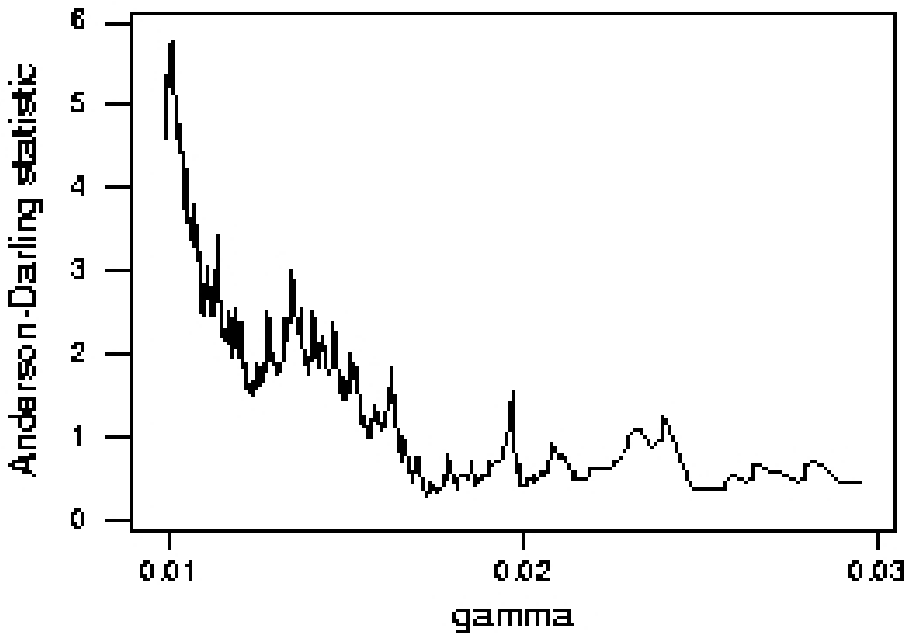}}
\caption{Anderson-Darling $A^2$ statistic versus selected values of the
threshold parameter $\gamma$, corresponding to the positive values of the series $S(t)$, computed from the Dow Jones index data.The minimum value, 0.32, is
attained for $\gamma=0.0173$}
\label{maxa2djp}
\end{figure}

\begin{figure}[htb!]
\resizebox{0.50\textwidth}{!}{%
\includegraphics{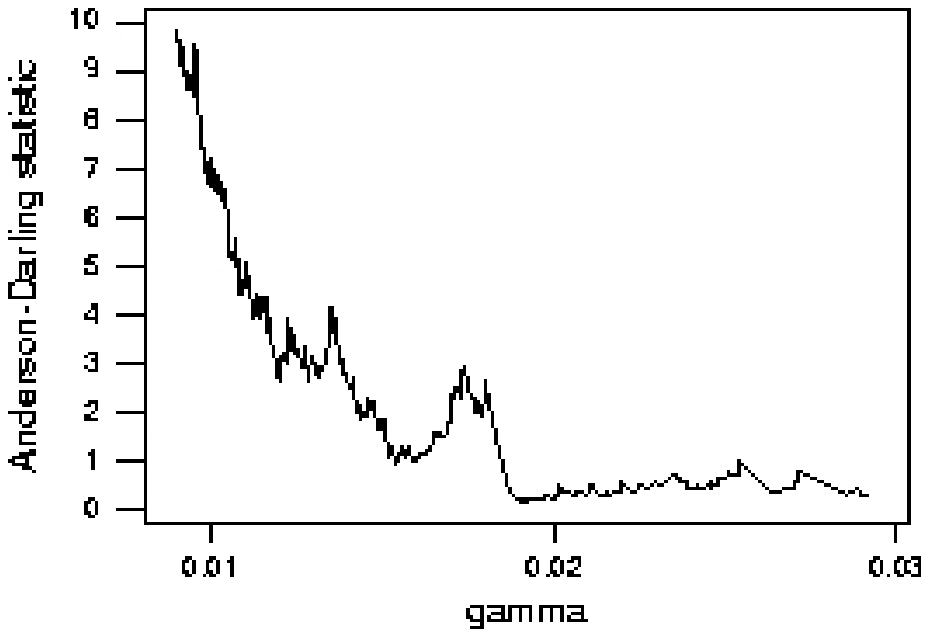}}
\caption{Anderson-Darling $A^2$ statistic versus selected values of the
threshold parameter $\gamma$, corresponding to the negative values of the series
$S(t)$, computed from the Dow Jones index data.The minimum value, 0.18, is
attained for $\gamma=0.0191$}
\label{maxa2djn}
\end{figure}

\noindent
Figures ~\ref{maxa2ipcp}
to ~\ref{maxa2djn} show the plots of $A^2$ 
versus different values of the threshold parameter $\gamma$. Using this approach, we obtained the following results:
\vspace{.08cm}

\noindent
For the IPC index data, the best possible fit for the largest values
in the positive tail, is obtained for $\gamma_0=0.0395$, where
the minimum value of $A^2$ is 0.16; based on the 64
largest positive observations, with an estimated value of
the shape parameter $\alpha=3.822$ For negative tail, the minimum value of $A^2$
was found to be 0.50, for $\gamma=0.036$. The fitted
value of $\alpha$ was $3.507$, based on the 69 smallest
observations. 

\noindent
For the case of the DJ index, the best positive fit gives
$A^2=0.315$ for $\gamma=0.0173$ with $\alpha=3.333$;  for
the negative tail the results showed that the smallest
value of $A^2=0.18$ is attained for $\gamma=0.0191$
with $\alpha=3.495$; the above results were obtained from
the 158 largest and the 124 smallest observations of $S_{DJ}(t)$,
respectively.

\noindent
Table ~\ref{results} summarize these results for easy reference.

\begin{table}
\begin{centering}
\begin{tabular}{|c|c|c|c|c|c|}
\hline
Index&Tail& $A^{2}$& $\gamma$ & $\alpha$ & n \\
\hline
\hline
IPC& Positive& 0.16&0.0395&3.82&64 \\
IPC& Negative&0.50&-0.0360&3.51& 69 \\
DJ & Positive&0.32&0.0173&3.33&158 \\
DJ & Negative&0.18&-0.0191&3.50&124 \\
\hline
\hline
\end{tabular}
\caption{Best left and right tail fits, for the two analyzed series.}
\label{results}
\end{centering}
\end{table}


\begin{figure}[htbp]
\resizebox{0.50\textwidth}{!}{%
\includegraphics{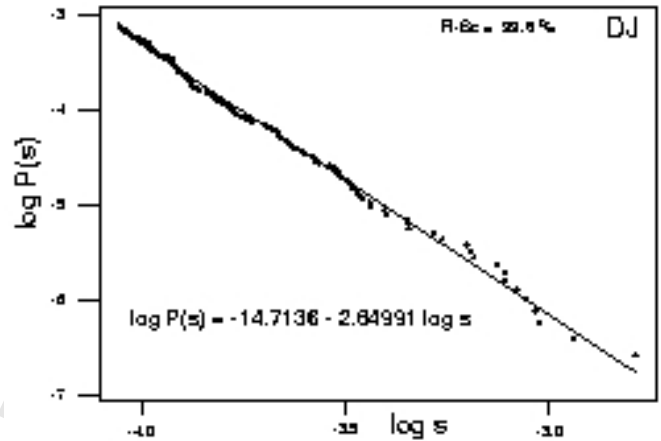}}
\caption{Linear fit for the positive tail of the Dow Jones $S(t)$ series
using $\gamma=0.0173$.The fitted value of the shape parameter $\alpha$ is 2.85}
\label{fit}
\end{figure}

\begin{figure}[htbp]
\resizebox{0.50\textwidth}{!}{%
\includegraphics{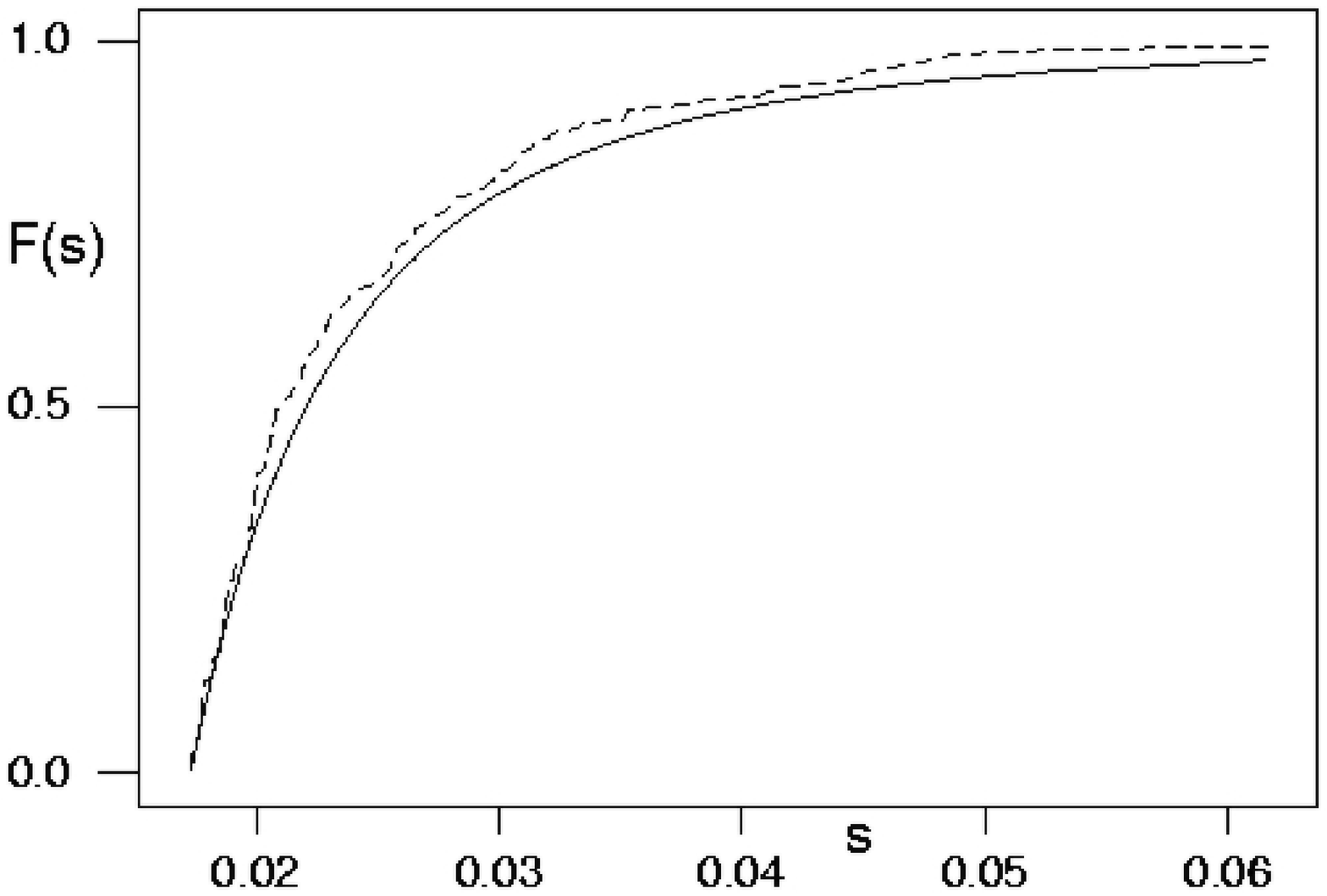}}
\caption{Empirical (dash) and fitted (solid) cumulative distribution functions
for the positive tail of the Dow Jones $S(t)$ series, using $\gamma=0.0173$ and $\alpha=2.85$}
\label{funipc}
\end{figure}

\begin{figure}[htbp]
\resizebox{0.50\textwidth}{!}{%
\includegraphics{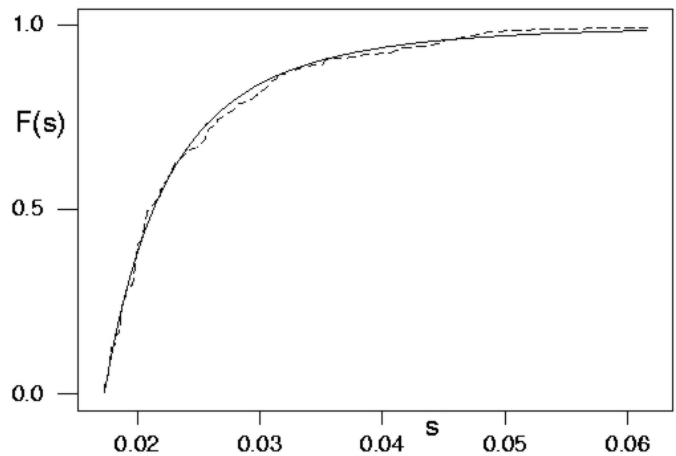}}
\caption{Empirical (dash) and fitted (solid) cumulative distribution functions
for the positive tail of the Dow Jones $S(t)$ series, using $\gamma=0.0173$ and $\alpha=3.33$}
\label{fundj}
\end{figure}

\begin{table}
\begin{centering}
\begin{tabular}{|c|c|c|c|}
\hline
Index&Tail& $A^{2}$-method& Graphical method \\
\hline
\hline
IPC& Positive& 3.82&3.00 \\
IPC& Negative& 3.51&2.82 \\
DJ & Positive& 3.33&2.85 \\
DJ & Negative& 3.50&2.80 \\
\hline
\hline
\end{tabular}
\caption{Fitted values of the parameter $\alpha$ obtained by graphical
and $A^{2}$ methods.}
\label{tablefit}
\end{centering}
\end{table}
\noindent
Table ~\ref{tablefit} suggests that our method tends to produce larger
fitted values for the parameter $\alpha$; however, we can expect a much better
fit using the parameter values produced by the $A^2$-method for obvious
reasons. As an illustration consider the fitted regression line of $\log P(s)$ on
$\log s$, where $P(s)=1-F_n(s)$, for the positive tail of the Dow Jones $S(t)$ series
shown in figure \ref{fit}. The fitted value of $\alpha=2.85$ corresponds to the slope
of the regression line and it differs from our estimate $\alpha=3.33$;  figures 
\ref{fundj} and \ref{funipc} show the empirical $F_n$ (dash), and the fitted cumulative
distribution function $F$ for each case, using $\gamma=0.0173$. As expected, our estimates
produce a better fit. 

\section{Conclusions}
\vspace*{-.3cm}

\noindent
An objective technique for fitting the Power-Law distribution to extreme variations
in stock market indexes was presented. The method is based on the use of
Anderson-Darling Statistic $A^2$ as a measure of discrepancy between the empirical and
the theoretical distribution functions, selecting as the fitted parameters, the values
which minimize such a measure. The technique was illustrated for the case of the
Dow Jones industrial average index and for the Mexican prices and quotations index.
The results showed that this method can be used with better results than the traditional
graphical method in which the value of the cutoff parameter $\gamma$ is chosen subjectively.

\begin{acknowledgement}
A.R.H.M wishes to thank Conacyt-Mexico for financial support provided under Grant No. 44598-F. Also we thank useful suggestions and bibliographical support from P. Giubellino.
\end{acknowledgement}
%
%

\end{document}